\documentclass[letter]{aa}  

\usepackage{graphicx}
\usepackage{txfonts}
\usepackage{mathabx}
\usepackage{hyperref}
\hypersetup{
    colorlinks=true,
    linkcolor=black,
    filecolor=black, 
    urlcolor=black, 
    citecolor=black, 
}

\newcommand{\orcid}[1]{\href{https://orcid.org/#1}{\includegraphics[width=8pt]{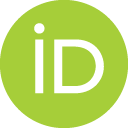}}}
\begin{document} 

   \title{A strong H$^-$ opacity signal in the near-infrared emission spectrum of the ultra-hot Jupiter KELT-9b}

   \author{Bob Jacobs
          \inst{1}\orcid{0000-0002-0373-1517}
          \and
          Jean-Michel D\'esert
          \inst{1}\orcid{0000-0002-0875-8401}
          \and
          Lorenzo Pino
          \inst{2}
          \and
          Michael R. Line
          \inst{3}
          \and 
          Jacob L. Bean
          \inst{4}
          \and
          Niloofar Khorshid
          \inst{1}
          \and
          Everett Schlawin
          \inst{5}
          \and
          Jacob Arcangeli
          \inst{1}
          \and
          Saugata Barat
          \inst{1}
          \and
          H. Jens Hoeijmakers
          \inst{6}
          \and 
          Thaddeus D. Komacek
          \inst{7}
          \and
          Megan Mansfield
          \inst{5}
          \and
          Vivien Parmentier
          \inst{8}
          \and
          Daniel Thorngren
          \inst{9}
          }

   \institute{Anton Pannekoek Institute for Astronomy, University of Amsterdam,
Science Park 904, 1098 XH,
Amsterdam, the Netherlands\\
              \email{b.jacobs@uva.nl}
         \and 
             INAF-Osservatorio Astrofisico di Arcetri, Largo Enrico Fermi 5 I-50125, Firenze, Italy
         \and
             School of Earth \& Space Exploration, Arizona State University, Tempe, AZ 85257, USA
        \and
             Department of Astronomy and Astrophysics, University of Chicago, Chicago, IL 60637, USA
        \and
             Steward Observatory, The University of Arizona, 933 North Cherry Avenue
Tucson, AZ 85721, USA
        \and
            Lund Observatory, Department of Astronomy and Theoretical Physics, Lund University, Box 43, 221 00 Lund, Sweden
        \and
            Department of Astronomy, University of Maryland, College Park, MD 20742, USA
        \and
             Department of Physics (Atmospheric, Oceanic and Planetary Physics), University of Oxford, Oxford, OX1 3PU, UK
        \and
             Institute for Research on Exoplanets (iREx), Université de Montréal, Montreal, QC, Canada
             }

   \date{Received July 18, 2022 / Accepted November 13, 2022}

  \abstract
   {We present the analysis of a spectroscopic secondary eclipse of the hottest transiting exoplanet detected to date, KELT-9b, obtained with the Wide Field Camera 3 aboard the \textit{Hubble Space Telescope}. 
   We complement these data with literature information on stellar pulsations and \textit{Spitzer}/Infrared Array Camera and \textit{Transiting Exoplanet Survey Satellite} eclipse depths of this target to obtain a broadband thermal emission spectrum.
   
   Our extracted spectrum exhibits a clear turnoff at $1.4$\,$\mu$m. This points to H$^{-}$ bound-free opacities shaping the spectrum.
   To interpret the spectrum, we perform grid retrievals of self-consistent 1D equilibrium chemistry forward models, varying the composition and energy budget.

   The model with solar metallicity and C/O ratio provides a poor fit because the H$^{-}$ signal is stronger than expected, requiring an excess of electrons. This pushes our retrievals toward high atmospheric metallicities  ($[M/H]=1.98^{+0.19}_{-0.21}$) and a C/O ratio  that is subsolar by 2.4$\sigma$. We question the viability of forming such a high-metallicity planet, and therefore provide other scenarios to increase the electron density in this atmosphere.
   We also look at an alternative model in which we quench TiO and VO. This fit results in an atmosphere with a slightly subsolar metallicity and subsolar C/O ratio ($[M/H]=-0.22^{+0.17}_{-0.13}$, log(C/O)$=-0.34^{+0.19}_{-0.34}$). However, the required TiO abundances are disputed by recent high-resolution measurements of the same planet.

   }

   \authorrunning{Bob Jacobs et al.}
   \maketitle
%

\section{Introduction}

Ultra-hot Jupiters (UHJs) are a class of exoplanets ($T_{eq} \gtrsim 2200$\,K) that are distinct from hot Jupiters ($1000$\,K $\lesssim T_{eq} \lesssim 2200$\,K). At temperatures greater than 2200\,K most molecules become partially thermally dissociated and alkalies are mostly ionized \citep{Lothringer2018,Parmentier2018}. On the dayside H$_2$ is partially dissociated, and as winds take the hydrogen atoms to the terminator, they recombine. This exothermic reaction transports energy efficiently to the nightside \citep{BellCowan2018, Mansfield2020}.

The dissociated hydrogen can interact with electrons that are freed by the thermal ionization of metals with low ionization energies like Na, K, and Fe, through free-free or bound-free absorption of a photon. Both can provide significant continuum opacities at infrared wavelengths that dominate UHJ emission spectra and mute H$_2$O features typical for clear  hot-Jupiter atmospheres \citep{Arcangeli2018}. However, the bound-free opacity drops off sharply above 1.4\,$\mu$m \citep{Lenzuni1991}, which could therefore be detectable with the G141 grism ($1.1-1.7$\,$\mu$m) of the \textit{Hubble Space Telescope} Wide Field Camera 3 (\textit{HST}/WFC3). UHJs with measured near-infrared emission are likely not hot enough to directly detect this H$^{-}$ feature since their spectra exhibit H$_2$O features, thus masking the H$^{-}$ \citep{Arcangeli2018}.

KELT-9b \citep{Gaudi2017}, a 2.88 $M_J$ planet that orbits an A0 host star on a 1.48 day orbit, opened up this opportunity as it is by far the hottest UHJ currently known. With a dayside brightness temperature of 4566\,K at $4.5$\,$\mu$m \citep{Mansfield2020}, the KELT-9b  atmosphere is likely in thermochemical equilibrium \citep{Kitzmann2018, Parmentier2018}. \citet{Pino2020} showed that the dayside atmosphere has a thermal inversion that is likely driven by atomic metals and metal hydrides and the lack of atmospheric cooling due to thermal molecular dissociation.

Recently, \citet{Wong2021b} and \citet{Mansfield2020} analyzed KELT-9b phase curves at $0.8$\,$\mu$m and $4.5$\,$\mu$m with TESS and Spitzer, respectively. Their interpretations require H$_2$ dissociation and recombination to explain the low day-night contrast. Their phase curves contained stellar pulsations with a period of 7.59 hours and peak-to-peak amplitudes of $260\pm12$ ppm and $280\pm44$ ppm, respectively. These $\delta$ Scuti-type pulsations were spectroscopically confirmed by \cite{Wyttenbach2020}.

\citet{Changeat2021} analyzed the \textit{HST}/WFC3 observations of the near-infrared emission spectrum of KELT-9b that is   presented in the current study. They used a free-retrieval technique to retrieve individual molecular abundances for a parametric temperature-pressure profile. \citet{Changeat2021} obtained an atmosphere with high abundances of TiO, VO, and FeH, molecules that should be dissociated in this atmosphere. They argued that hydrodynamic escape from the top of the atmosphere \citep{Wyttenbach2020} lofts these molecules from deeper atmospheric layers where the temperature is low enough for them to exist in equilibrium. Using the same data, we present a different data reduction and analysis resulting in a different emission spectrum. We also use a grid-based atmospheric retrieval of self-consistent models. Consequently, we obtain an alternative interpretation of the  KELT-9b emission spectrum that does not require TiO, VO, and FeH.

\section{Data analysis}
\label{sec:obs}
\subsection{Observations}
We observed one secondary eclipse of KELT-9b with six orbits with WFC3 for Program GO 15820 (PI: L. Pino). The data were obtained with the G141 grism, covering 1.1 to 1.7\,$\mu$m, using the bi-directional spatial scanning technique. 

\subsection{Data reduction}
We based our custom data reduction pipeline on the pipeline developed by \citet{Arcangeli2018}, which forms subexposures from each full exposure by subtracting consecutive  nondestructive reads.
We made some small modifications to improve the pipeline. We calibrated the wavelength solution by matching the first exposure in eclipse to a convolution of a PHOENIX stellar model \citep{Husser2013} for KELT-9 (with $T_{\rm{eff}}=10,200$\,K) with a transmission profile of G141 following the method outlined in \citet{Wilkins2014}. 
We identified cosmic rays using a local median filter, clipping pixels that are five median deviations from the median.
\citet{Arcangeli2018} took the standard deviation above the mean, which is more prone to missing cosmic rays that affect multiple pixels. We also flagged cosmic ray affected pixels in the next subexposure because of their persistence.

Subsequently we performed an optimal extraction algorithm \citep{Horne1986} on every subexposure to maximize the signal-to-noise ratio. To combine the spectra we shifted and shrank the resulting spectra of each subexposure by at most 1.5 pixels and 1.3\%, respectively, to match the wavelength grid of the first subexposure of the visit. This negates the warp of the G141 grism response on the detector \citep{Tsiaras2016} without the need to split (bad) pixels.

\subsection{Systematics correction}
WFC3 light curves are dominated by time-dependent systematics   \citep[e.g.,][]{Deming2013}. They are strongest in the first orbit, which we therefore discard. To remove the systematics in the other orbits we employed two different techniques: the \texttt{RECTE} charge trap model developed by \citet{Zhou2017} and a \texttt{divide-in-eclipse} model similar to the divide-out-of-transit method \citep{Berta2012}. 

In the \texttt{RECTE} method we use four parameters to model the orbit-ramps: the number of initially filled fast or slow charge traps ($E_{f/s,0}$) and the number of fast or slow charge traps that are filled in between orbits ($\Delta E_{f/s}$). These parameters are the same for each scan direction.

\begin{figure}
    \centering
    \includegraphics[width=\columnwidth]{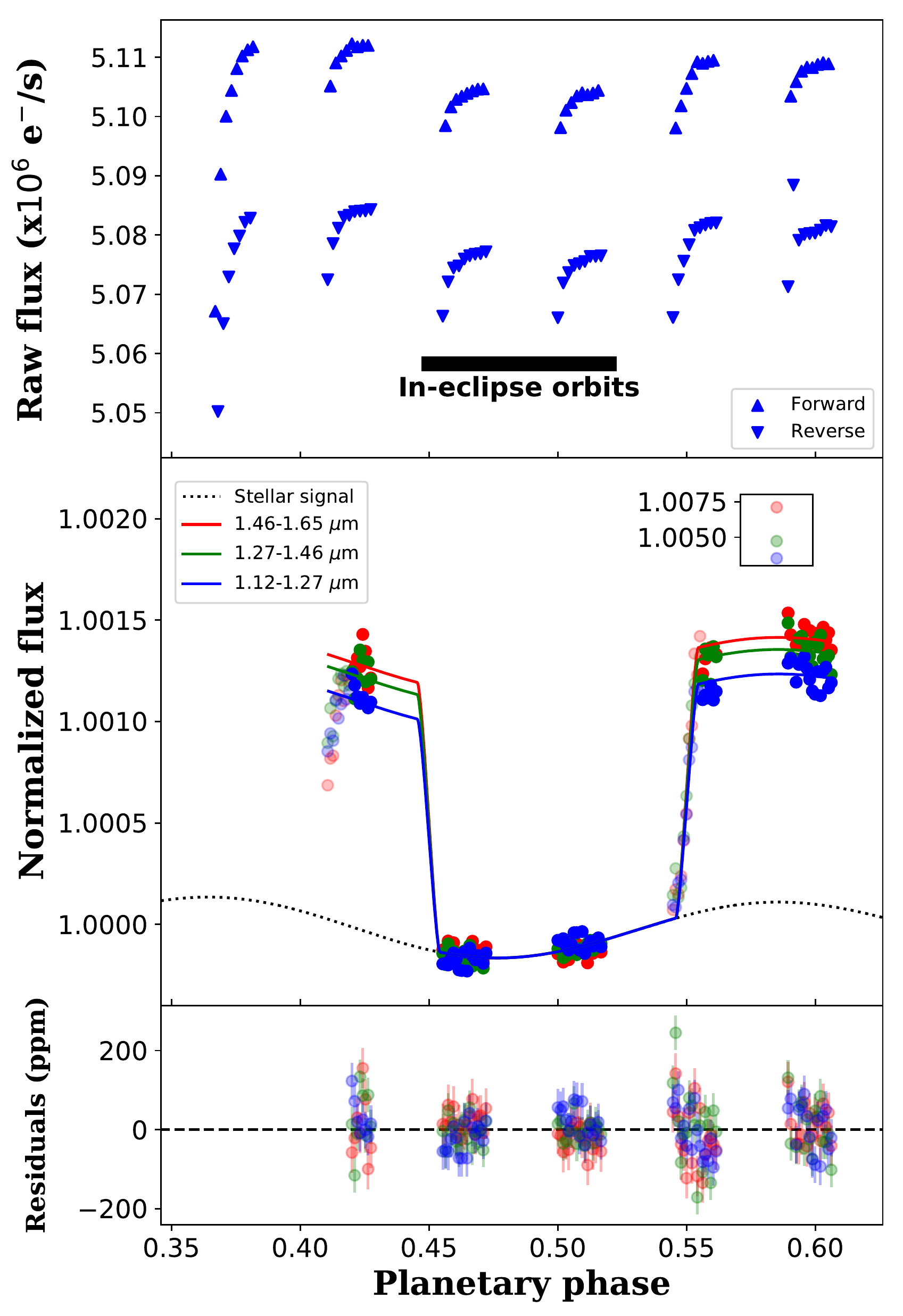}
    \caption{Eclipse light curve of KELT-9b.\\\textbf{Upper:} Raw light curve of \textit{HST}/WFC3 for wavelengths of $1.12-1.27$\,$\mu$m for forward and reverse scanned data. We divide each orbit by the average of the two in-eclipse orbits. The result of this \texttt{divide-in-eclipse} method is shown in the middle panel in blue. \\
    \textbf{Middle:} Eclipse light curve of \textit{HST}/WFC3 split into three wavelength bins. The solid lines are best fits; the dotted line shows the stellar signal including pulsations, ellipsoidal variations, and Doppler boosting.  The first orbit was discarded entirely. The inset shows the exposure affected by a satellite crossing. In our fits  the grayed-out points in the first orbit, the points in egress, and the satellite crossing were excluded. \\
    \textbf{Lower:} Residuals from the fit in the middle panel. The \texttt{divide-in-eclipse} method induces larger residuals outside eclipse, although a deviation from our model is still visible in egress.\\ The \texttt{RECTE} equivalent of the lower two panels is shown in Appendix \ref{sec:apprecte}.}
    \label{fig:dataplots}
\end{figure}

In the \texttt{divide-in-eclipse} method we divide the wavelength-dependent light curves by the mean of the two in-eclipse orbits (see Figure \ref{fig:dataplots}). This leaves a ramp profile in the first nine exposures of the second orbit, which we therefore discard as well. Discarding the first eight or ten exposures of the second orbit changes the spectrum by \textless$0.5\sigma$.
Fitting this single ramp with an exponential model like \citet{Kreidberg2014a} proved to be difficult for smaller wavelength bins because of a combination of a smaller ramp amplitude and larger scatter; it increased eclipse depth error bars by a factor of 2. 

Additional fit parameters shared between the two  methods are the eclipse depth and linear visit-long slopes for forward scanned and   reverse scanned data. As done in \citet{Wong2021b}, we multiplied the stellar signal by an ellipsoidal variation and a Doppler boosting signal with theoretical amplitudes of 38\,ppm and 1.6\,ppm, respectively. We modeled the eclipse light curve with \texttt{batman} \citep{Kreidberg2015}, fixing the remaining system parameters to literature values from \citet{Wong2021b}.

We modeled  the stellar pulsations with a pulsation phase parameter and we fixed its period and monochromatic peak-to-peak amplitude to the values found by \citet{Wong2021b}: 7.59 hours and $260$\,ppm, respectively. We find a model with a polynomial baseline and a sine baseline with the planet's period to be disfavored by BIC, but a model including the stellar pulsations to be favored by a BIC factor of 58 for \texttt{divide-in-eclipse} and 31 for \texttt{RECTE}. 
Furthermore, if we split the spectrum into three wavelength bins we consistently find the same pulsation phase across both methods and all wavelength bins.
The inclusion of pulsations decreases eclipse depths by 135\,ppm monochromatically corresponding to a 176\,K drop in brightness temperature. This is over four times our spectroscopic precision of 39\,ppm.
Changing the pulsation phase by 3$\sigma$ changes the spectrum by either +41 or -31\,ppm.

The third exposure of the last orbit was affected by a satellite-crossing event \citep[see][]{Fu2020}. We found no signs of its persistence impacting the remainder of the orbit, as  in \citet{Fu2020}. We therefore simply discarded this exposure. 

Because we expected large variations in eclipse depths over the spectrum, we split the white light curve into three bins and fit to find the stellar pulsation phase. We plotted the light curve for the \texttt{divide-in-eclipse} method in Figure \ref{fig:dataplots}.
In these fits we excluded all but the last six exposures of the fourth orbit because an isotropic disk-integrated planetary model does not provide a good fit to the data in egress (see Appendix \ref{sec:app_egress}). Finally, we split the spectrum into 23  five-pixel wide bins and reach an average spectrophotometric precision of 10\% for \texttt{divide-in-eclipse} and 7\% above photon noise for the \texttt{RECTE} method.

\subsection{Estimation of errors}
In order to estimate the errors on our fitted parameters and identify the degeneracies in the model we adopted a Markov chain Monte Carlo (MCMC) approach using the open-source \texttt{emcee} code \citep{emcee}. 
Each chain has 10,000 steps with 100 walkers.
Our average final precision on the spectroscopic eclipse depths with \texttt{divide-in-eclipse} and \texttt{RECTE} are 39 ppm and 38 ppm, respectively, per wavelength bin for 23 bins.

\begin{figure*}
    \centering
    \includegraphics[width=\textwidth]{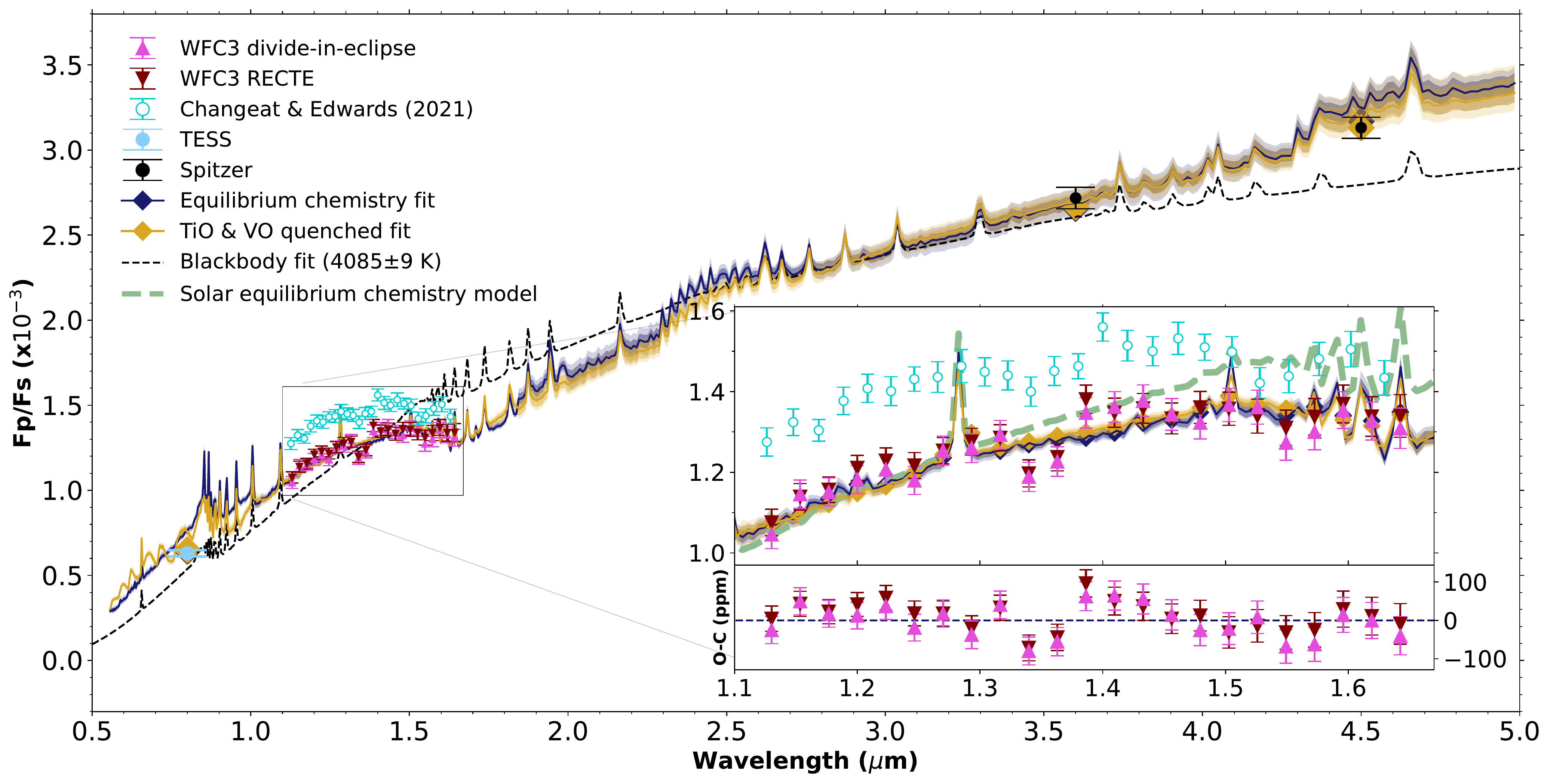}
    \caption{Dayside eclipse spectrum from this work (red--pink), \citet{Changeat2021} (turquoise) and previous TESS/Spitzer observations (light blue--black) compared to the best-fit model spectra for equilibrium chemistry (dark blue), TiO and VO quenched chemistry (yellow), and a blackbody (dashed black). The shaded regions denote 1$\sigma$ and 2$\sigma$ model intervals. Spikes in the blackbody spectrum are stellar absorption lines. The inclusion of stellar pulsations causes an offset between the \citet{Changeat2021} data and our spectrum. In the inset we zoom in on the WFC3 spectrum and show the residuals of the data with respect to the equilibrium chemistry fit. The dashed green line shows a solar equilibrium chemistry model. This green model already deviates from the blackbody model redward of 1.4\,$\mu$m due to H$^{-}$ opacities. The KELT-9b H$^{-}$ turnover is even larger than for this solar model.}
    \label{fig:spectrum}
\end{figure*}

\section{Results}
\label{sec:results}

We show our WFC3 emission spectra in red and pink in Figure \ref{fig:spectrum}. We complement them with a TESS eclipse depth \citep{Wong2021b}, a 4.5\,$\mu$m Spitzer eclipse depth \citep{Mansfield2020}, and a 3.6\,$\mu$m Spitzer eclipse depth (Beatty et al. in prep.). The spectra have a distinct turnoff at 1.4\,$\mu$m, away from a blackbody model, that we attribute to H$^{-}$ bound-free opacities. This strongly supports previous studies of UHJs that required H$^{-}$ opacities to fit their spectra \citep[e.g.,][]{Arcangeli2018, Mikal-Evans2019}. The turnoff is weaker than in the spectrum of \citet{Changeat2021}.
This is confirmed by the \texttt{RECTE} spectrum, which is consistent with the \texttt{divide-in-eclipse} spectrum up to $0.5\sigma$ with the exception of a 12 ppm offset that is due to a 1$\sigma$ difference in pulsation phase. Because of the similarity we   therefore performed our retrievals on the \texttt{divide-in-eclipse} spectrum only.

\subsection{Equilibrium chemistry}
\begin{figure*}
    \centering
    \includegraphics[width=\textwidth]{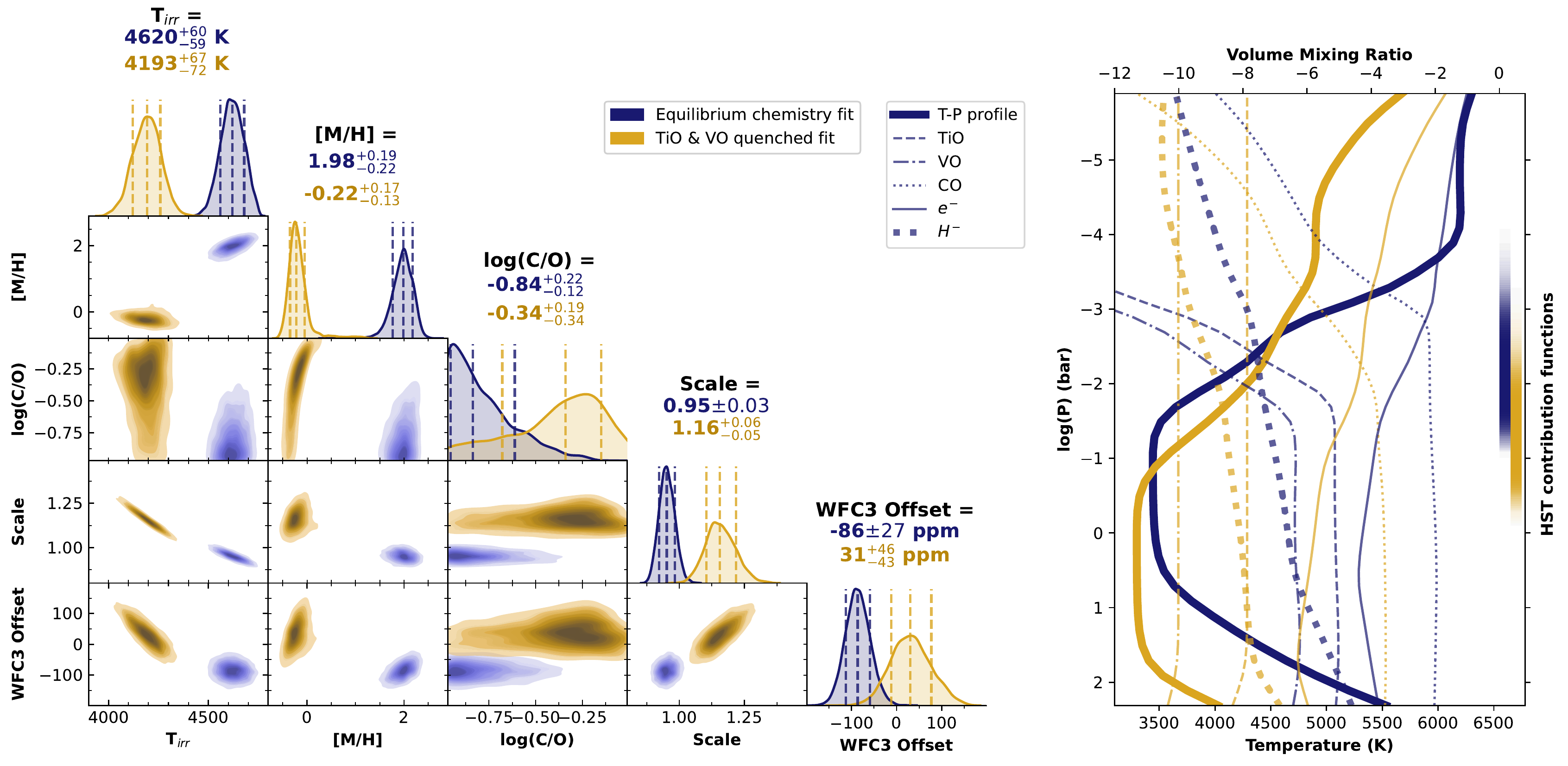}
    \caption{\textbf{Results from the grid retrievals. Left:} Posterior distributions from the grid retrievals for the equilibrium chemistry model (dark blue) and the TiO--VO quenched model (yellow). \textbf{Right:} Temperature-pressure profiles and the volume mixing ratios of a selection of the most important species for  the equilibrium chemistry model and the TiO--VO quenched model. In the TiO--VO quenched model the TiO and VO VMRs are artificially enhanced to the maximum value of the atmospheric layers below, which is at 20\,bar. Also plotted are the HST contribution functions. Spitzer probes the same pressure regions as HST.}
    \label{fig:mcmc}
\end{figure*}

We performed parameter estimation over an ScCHIMERA (self-consistent CHIMERA) grid of thermochemical equilibrium chemistry forward models (see Appendix \ref{sec:appgridmodels})
using \texttt{emcee} to obtain the best fitting irradiation temperature, metallicity, C/O ratio, scale factor, and WFC3 offset. We show the results in dark blue in Figures \ref{fig:spectrum} and \ref{fig:mcmc}. We achieve a reduced chi-squared statistic of 1.41, much better than for a blackbody fit with $\chi_{\nu} = 14$.
In the absence of H$_2$O, the C/O ratio is largely insensitive to the WFC3 spectrum, and is instead driven by the $4.5$\,$\mu$m Spitzer point, which covers the CO emission feature. This drives the fit toward the lower log(C/O) limit of $-1$ for a C/O ratio that is subsolar by 2.4$\sigma$.

The H$^{-}$ bound-free opacity is higher than expected for a solar metallicity and C/O ratio equilibrium atmosphere model (green in Figure \ref{fig:spectrum}). An increased production of H$^{-}$ requires an enhancement of the free electron density. Nearly all K, Na, Ca, and Fe atoms are singly thermally ionized at the temperatures and pressure of the  KELT-9b WFC3 photosphere. The temperature is too low to thermally ionize any elements twice. This forces our fits into a significantly supersolar metallicity solution of $[M/H]=1.98^{+0.19}_{-0.22}$.

The retrieval finds that if our WFC3 eclipse depths were increased by $86\pm27$\,ppm, they would fit the TESS and Spitzer data better. However, any such offset does not alter the metallicity or the C/O ratio significantly. Possible sources of this offset include instrumental effects, uncertainties in stellar pulsations, differences between epochs, and differences between a secondary eclipse measurement and a phase curve measurement. We tested the difference between a linear planetary emission baseline and a phase curve by modeling the phase curve with a sine with a fixed nightside brightness temperature of 2566\,K and a phase offset of 18.7$^{\circ}$ (analogous to the findings of \citealt{Mansfield2020}). This resulted in a mean WFC3 eclipse depth increase of 45\,ppm, but also made the 1.4\,$\mu m$ turnover slightly sharper, requiring more bound-free absorption, and therefore even more electrons, and in turn more metals. This increase is smaller for higher nightside temperatures and smaller phase offsets, as found by \citet{Wong2021b}. 

\subsection{Quenched TiO--VO model and comparison with previous studies}
There is some structure in the residuals between the data and the model spectrum around 1.4\,$\mu$m. 
We inspect the authenticity of this feature in Appendix \ref{sec:135feature}. These points likely drove the free retrieval of \citet{Changeat2021} toward a solution with high amounts of TiO and especially VO.

To explore the effects of the lofting of elements from deep atmospheric layers, as suggested by \citet{Changeat2021}, we modified our self-consistent equilibrium models by quenching TiO and VO \citep[see, e.g.,][]{Tsai2017} by freezing TiO and VO volume mixing ratios (VMRs) in the upper atmosphere to the maximum VMR of the layers below. This effectively enhances TiO and VO in the WFC3 photosphere by a few orders of magnitude; however, it keeps their relative abundance tied to each other, and to  the H$^{-}$ abundance, through the metallicity parameter. We show the resulting fitted spectrum and atmospheric properties in yellow in Figures \ref{fig:spectrum} and \ref{fig:mcmc}. 

The retrieved metallicity of $[M/H]=-0.22^{+0.17}_{-0.13}$ and C/O ratio of $0.46\pm0.25$ are closer to the expected solar values, but the residual feature around 1.4\,$\mu$m is still not fit. With a $\chi_\nu^2=1.57$ this fit is also marginally worse than the equilibrium model fit.

\section{Discussion}
\label{sec:discussion}

\subsection{This work in the broader context of the KELT-9b studies}
We give a detailed analysis of the differences between the present study and \citet{Changeat2021} in Appendix \ref{sec:changeat}.

The metallicity-planetary mass trend tentatively seen in exoplanet atmospheres \citep{Kreidberg2014b, Arcangeli2018, Welbanks2019} predicts that a 2.88$M_J$ planet like KELT-9b   most likely has an atmospheric metallicity similar to its host star's metallicity. Given the host star's metallicity of [Fe/H] $=0.14$ \citep{Borsa2019}, a $\sim$$100\times$ solar metallicity atmosphere would significantly deviate from this trend.

As pointed out by \citet{Taylor2020}, 1D retrievals can be biased due to the inherent 2D nature of the planet's emission. To best account for this effect we  included their scale factor in our retrievals. At a nightside temperature of 2500--3000\,K \citep{Wong2021b, Mansfield2020}, the limbs of  KELT-9b  could host molecules like TiO, VO, and FeH. However, because most of the KELT-9b flux should come from the hotter substellar point, it is unlikely that the molecules in the limbs can induce the residual feature at 1.4\,$\mu$m that even a dayside-wide TiO--VO quenched model is unable to replicate. 

Recently, \citet{Kasper2021} observed the KELT-9b emission in high resolution. They place an upper limit to the TiO VMR of $10^{-8.5}$ at 99\% confidence. This contradicts the \citet{Changeat2021} atmospheric models as well as our best-fit TiO--VO quenched model, which has a TiO VMR of $10^{-7.9}$. 

\citet{Pino2020} find a supersolar Fe\,\textsc{I} abundance in the KELT-9b emission spectrum at 2$\sigma$, although their analysis was marginalized over a single assumed temperature profile, which is degenerate with iron abundance, and therefore does not provide a comprehensive analysis of the abundance. \citet{Borsa2022} observed neutral oxygen lines that are consistent with a solar abundance. However, we note that the total atmospheric oxygen abundance also includes the ionized species. Even though \citet{Kasper2021} retrieve a solar Fe \textsc{I} abundance, their retrieval requires an Fe\,\textsc{II} abundance that is 100 times higher. Moreover, even though the  \citet{Hoeijmakers2019} Fe\,\textsc{I} and Sc\,\textsc{II} transit depths are nicely explained by a solar metallicity 4000\,K model, they find that transit depths of other species, most notably Fe\,\textsc{II}, are deeper.
While highly dependent on the assumed and retrieved temperature, these increased ionized iron abundances provide support for our equilibrium chemistry model that also requires more ionized metals than at solar abundance to fuel H$^{-}$ opacities with free electrons. Because \citet{Kasper2021} probe a higher altitude of the atmosphere, their solar H$^{-}$ abundance and TiO VMR upper limit $10^{-8.5}$ are still compatible with our high-metallicity model. 

\subsection{Explaining the strong H$^{-}$ signal}
In Appendix \ref{sec:app_niloo} we simulate the formation of a KELT-9b-like planet through planetesimal accretion during type II migration, and we show that our high-metallicity case requires additional ways to enrich the KELT-9b atmosphere with metals. The solid core could be dissolved into the atmosphere \citep{Madhu2017, Bitsch2018}. A $100\times$ solar atmospheric metallicity is consistent at only 2$\sigma$ of the fully mixed dissolved core metallicity prior upper limit from \citet{Thorngren2019}. However, that model excludes planetary mass loss and its priors are based purely on the masses of warm Jupiters, while KELT-9b's insolation is a strong outlier. Alternatively, planetesimals that could be accreted in situ post-formation could have higher abundances of the primary electron donors relative to other refractories \citep{Schneider2021} or volatiles \citep{Welbanks2019}. By combining some of the above scenarios a $100\times$ solar metallicity could be possible, but it is improbable. A high metallicity due to the loss of the  KELT-9b initial envelope is unlikely because its mass loss rate is currently less than $10^9$\,kg\,s$^{-1}$ \citep{Borsa2022}. Moreover, this rate already facilitates the escape of Fe and O. 

The metallicity parameter in our equilibrium chemistry fit is effectively a proxy for the electron density that is required for the strong H$^{-}$ absorption in the WFC3 photosphere. The atmosphere of KELT-9b  may therefore have a higher ionization fraction than computed by ScCHIMERA, rather than a higher metallicity. Most UHJ spectra in \citet{Mansfield2021} might also show a stronger H$^{-}$ signal than they model with ScCHIMERA. In the KELT-9b WFC3 photosphere Na, K, and even Fe are almost fully singly-ionized and hydrogen is starting to get ionized. 
Photoionization could propel the ionization process and increase the electron density by ionizing thermally excited atoms. This is motivation to use ion chemistry in current kinetics models to include the combination of disequilibrium via mixing and photochemistry. \citet{Fossati2020, Fossati2021} show that inaccurate H$^{-}$ abundance modeling or nonlocal thermal equilibrium (NLTE) effects can also increase the H$^{-}$ abundance, for example through an increased ionization fraction. NLTE effects could also amplify the thermal inversion \citep[see Figure 3 in][]{Fossati2021}, which would have the consequence of increasing the H$^{-}$ turnoff in the emission spectrum.

\begin{acknowledgements}
      J.M.D acknowledges support from the Amsterdam Academic Alliance (AAA) Program, and the European Research Council (ERC) European Union’s Horizon 2020 research and innovation program (grant agreement no. 679633; Exo-Atmos). This work is part of the research program VIDI New Frontiers in Exoplanetary Climatology with project number 614.001.601, which is (partly) financed by the Dutch Research Council (NWO). Support for program HST-GO-15820 was provided by NASA through a grant from the Space Telescope Science Institute, which is operated by the Associations of Universities for Research in Astronomy, Incorporated, under NASA contract NAS5-26555. M.M. acknowledges support from NASA through NASA Hubble Fellowship grant HST-HF2-51485.001-A awarded by the Space Telescope Science Institute.
\end{acknowledgements}

\bibliographystyle{aa}
\bibliography{references}

\appendix

\section{The \texttt{RECTE} method light curve}
\label{sec:apprecte}
\begin{figure}
    \centering
    \includegraphics[width=\columnwidth]{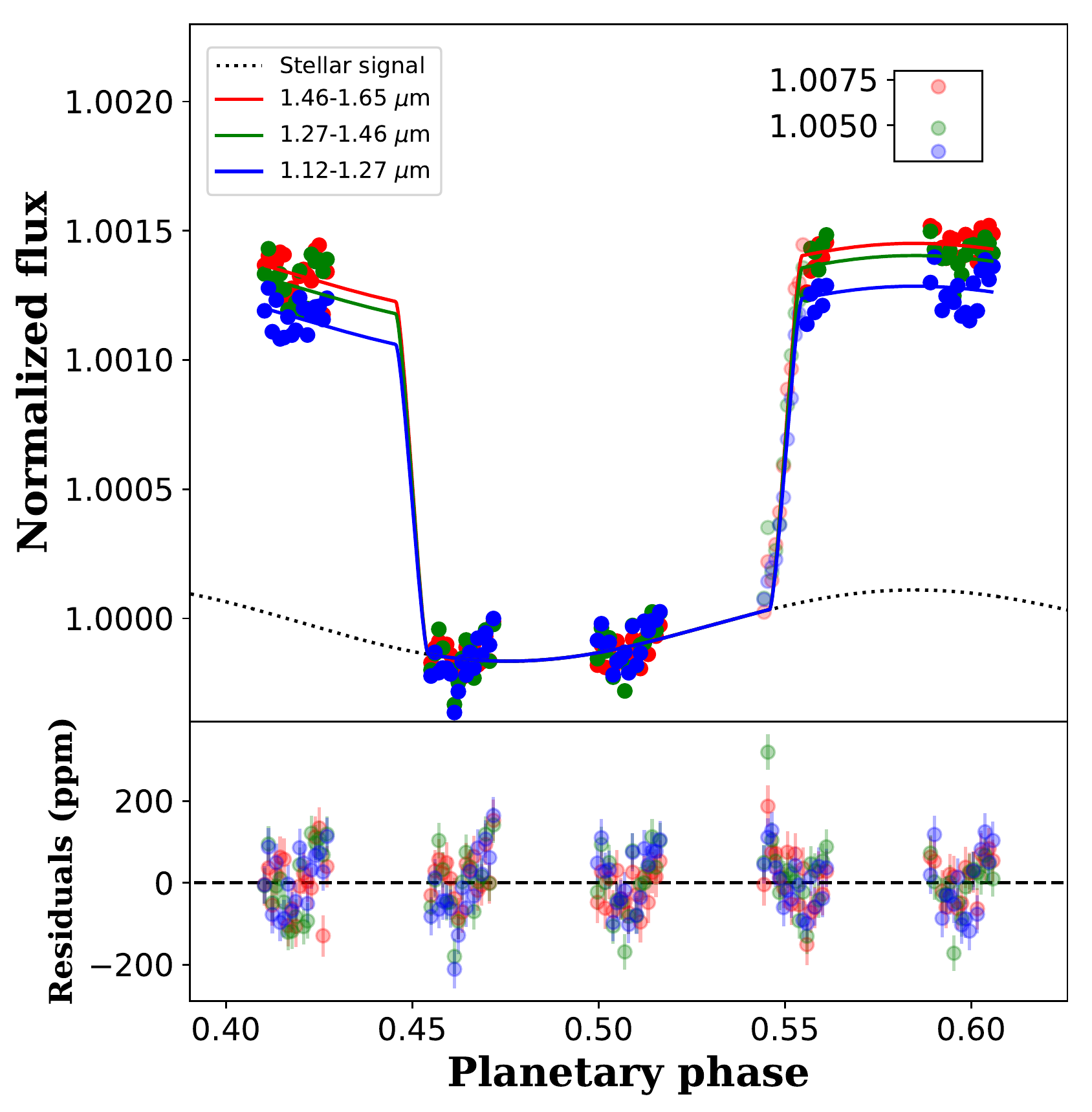}
    \caption{Same  as the lower two panels of Figure \ref{fig:dataplots}, but for the \texttt{RECTE} method.}
    \label{fig:dataplots-recte}
\end{figure}

In Figure \ref{fig:dataplots-recte} we show the three light curves when telescope systematics are corrected with the \texttt{RECTE} method. Significant correlated noise is left in the residuals. \texttt{RECTE} is unable to capture the shape of the orbit-ramp correctly, particularily at bluer wavelengths, because of the wide ranges of flux levels within the three wavelength bins and the relatively high flux in the blue wavelength bin. The ratio of the maximum to minimum flux of KELT-9 in the G141 transmission range is approximately 2. Therefore, there is a wide range in charge trap levels within one color bin, and the \texttt{RECTE} method is unable to capture this successfully. The correlated noise is significantly reduced for the 23 smaller spectral bins.

\section{Eclipse mapping egress}
\label{sec:app_egress}
Our isotropic disk-integrated planetary model results in significant residual structure in egress (see Figures \ref{fig:dataplots} and \ref{fig:dataplots-recte}). 
Changing orbital parameters to 3$\sigma$ away from their literature values cannot stretch egress enough to erase the deviation. 

We qualitatively assess these residuals by investigating the non-isotropicity of KELT-9b's dayside and a possible hotspot offset. We substitute our isotropic model for the   double-gray General Circulation Model (GCM) from \citet{TanKomacek2019} with three different drag timescales: $\tau_d=10^3, 10^5$, and $10^7$ s. These models have hotspots with intrinsic offsets of 0.2, 2, and 5 degrees, respectively. We fit light curves of these models to our \texttt{divide-in-eclipse} WFC3 light curve for the same three wavelength bins as in Figure \ref{fig:dataplots}. The free parameters are the planetary flux, a telescope systematics baseline, and a rotation of the GCM map. This rotation plus the model's intrinsic hotspot offset is the effective hotspot offset that we would like to find. We assume the same stellar pulsation amplitude and phase used for the isotropic model fit.

For the red, green, and blue bins we find hotspots offsets of $12\pm2^{\circ}$, $17\pm 2^{\circ}$, and  $22 \pm 3^{\circ}$, respectively. These offsets are similar to the offset of $18.7$\textdegree$ ^{+2.1^{\circ}}_{-2.3^{\circ}}$ found by \citet{Mansfield2020} in the Spitzer $4.5\mu$m phase curve and larger than the offset of $2.6$\textdegree$ ^{+1.4^{\circ}}_{-1.3^{\circ}}$ found by \citet{Wong2021b} in the TESS phase curve. The hotspot offsets appears to be insensitive to $\tau_d$ in our fits. A note of caution is necessary here: as this fit is sensitive to just a few data points, it is sensitive to small variations in data reduction and even stellar pulsation amplitude. Notably, we consistently find smaller hotspot offsets in the longer wavelength WFC3 bin than in the shorter wavelength bin, while the hotspot offset in the shorter wavelength TESS bandpass is smaller than in the longer wavelength Spitzer $4.5\mu$m bandpass. The larger offset in the Spitzer phase curve could also be explained by gravitational darkening induced seasons \citep{Wong2021b}.

To achieve better constraints on the hotspot offset in the WFC3 bandpass we would also require data in ingress and measurements with JWST or Cheops.

\section{Grid of atmospheric models}
\label{sec:appgridmodels}
To retrieve the  KELT-9b atmospheric properties, we generate a grid of self-consistent radiative-convective thermochemical equilibrium 1D models with ScCHIMERA \citep[self-consistent CHIMERA][]{Line2013}, which was validated against brown dwarf models \citep{Marley2010} and
has subsequently been applied to UHJ datasets  \citep{Arcangeli2018, Mansfield2018, Kreidberg2018}. We opt to use this self-consistent grid-based approach as an alternative to the free retrievals used by \citet{Changeat2021}.

ScCHIMERA uses the Newton--Raphson iteration technique \citep{McKay1989} to minimize the vertical flux divergence between atmospheric layers. It solves the thermal radiative fluxes at each atmospheric layer using the \citet{Toon1989} two-stream source function technique and it solves the convective fluxes with mixing length theory prescribed in \citet{Hubeny2017}. The incident stellar flux is modeled with a PHOENIX stellar model \citep{Husser2013} attenuated at a disk-averaged airmass of $1/\sqrt{3}$. We assume an internal temperature of 200\,K. Molecular and ion abundances are derived using the NASA CEA2 Gibbs-free energy minimization routine \citep{Gordon1994} given the \citet{Lodders2009} solar abundances scaled by a metallicity, [$M/H$], parameter and a carbon-to-oxygen ratio, C/O, parameter. This approach includes vertically varying abundances from thermal dissociation. Only neutral and singly thermally ionized states are considered. Molecular, atomic, and ionic opacities are treated within the correlated-K ``resort-rebin'' framework \citep{Lacis1991, Amundsen2017} and includes H$^{-}$ bound-free \citep{John1988} and free-free \citep{Bell1987} absorption. 

In addition, our grid includes an irradiation temperature parameter, $T_\mathrm{irr}$, and a scale factor that accounts for uncertainties in stellar and planet parameters and for dilution due to dayside 3D effects, as  in \citet{Taylor2020}. A parameter fitting the offset between the WFC3 data and the TESS/Spitzer data is also included because heterogeneous datasets from different instruments are not necessarily compatible \citep[e.g.,][]{Fraine2014, Barstow2015}. We apply this offset to the model spectra between 1.0\,$\mu$m and 1.85\,$\mu$m and we limit it to a range between $-200$ ppm and $200$ ppm. 

\section{The 1.4\,$\mu$m residual feature}
\label{sec:135feature}
There is significant structure in the residuals between the data and the model spectrum around 1.4\,$\mu$m in Figure \ref{fig:spectrum}.

This feature shows up in three out of four subexposures, and there are no anomalies in fit parameters as a function of wavelength. There is no flat-field error or stellar line in the exposures at these wavelengths. We note the existence of a background star, $\sim$0.2\% as bright as KELT-9 that crosses the short-wavelength part of the spectrum of KELT-9. We can constrain the exposure extraction box in our pipeline such that it excludes the background star's flux. This does not remove the 1.4\,$\mu$m residual feature. 

Chromium hydride has a large cross section in the WFC3 G141 bandpass that aligns very well with the 1.4\,$\mu$m feature. While CrH has been tentatively found in the \textit{HST}/WFC3 spectrum of other hot Jupiters \citep{Braam2021}, it is likely to be dissociated in KELT-9b.

\section{Comparison to the previous study of \citet{Changeat2021}}
\label{sec:changeat}

In this appendix we detail some of the differences between this study and \citet{Changeat2021}. We compare the results for our WFC3+TESS+Spitzer fits to the results for their WFC3+TESS+Spitzer fits, which did not yet have the 3.6\,$\mu$m Spitzer point.

In the present study we   present a detailed light curve fit that results in a slightly different spectrum (see Figure \ref{fig:spectrum}).  \citet{Changeat2021} used an exponential ramp model, while we use a \texttt{divide-in-eclipse} method and a \texttt{RECTE} model to confirm the result. Contrary to \citet{Changeat2021}, we  include the first two exposures of every orbit into our analysis. We also observe the stellar pulsations as it is included in our framework. This is consistent with the stellar pulsations found in most other spectroscopic and photometric studies on KELT-9b \citep[e.g.,][]{Wyttenbach2020, Wong2021b}. Ignoring the pulsation has some effects on the emission spectrum. If we discard the first two exposures of every orbit like \citet{Changeat2021}, the stellar pulsations are still significant at a $\Delta$BIC of 14 for \texttt{divide-in-eclipse} and a $\Delta$BIC of 7 for \texttt{RECTE}. 
 
Another important difference between the  present work and \citet{Changeat2021} is that they used a free retrieval technique, whereas we use self-consistent models (see Appendix \ref{sec:appgridmodels}). \citet{Changeat2021} require large amounts of TiO and VO: log(TiO)$=-6.9^{+0.3}_{-0.4}$ and log(VO)$=-6.7^{+0.2}_{-0.2}$. 
However, we note that such molecules should be dissociated at KELT-9b temperatures, and therefore we look for alternative scenarios. Our self-consistent approach does not require a large abundance of TiO and VO; instead, we find that the spectrum can be explained by either a high metallicity ($[M/H] = 1.98^{+0.19}_{-0.21}$) or a solar metallicity ($[M/H]=-0.22^{+0.17}_{-0.13}$) with quenching of TiO and VO at abundances of log(TiO)$=-7.9$ and log(VO)$=-10.0$. We note that the relative abundance of titanium and vanadium is fixed in our self-consistent model. We find that the TiO and VO features in our model are also muted because of the presence of H$^{-}$ abundance that is almost an order of magnitude higher than in \citet{Changeat2021}. 

When fitting for the WFC3 spectrum only, \citet{Changeat2021} only require TiO and VO. When we fit our two atmospheric models (equilibrium chemistry and TiO--VO quenched fits) to the WFC3 data only, we find only marginally different metallicities and C/O ratios for the two models compared to fitting to the full spectrum.

\section{Planetesimal accretion simulation for KELT-9b}
\label{sec:app_niloo}
Using the planet formation code SimAb described in \citet{Khorshid2021}, we can predict what metallicities are possible for a planet with mass and orbital distance similar to KELT-9b, considering a simplistic planet formation scenario. SimAb simulates a planet core that goes through rapid gas accretion and is migrating inward through  type II migration. Planets in SimAb accrete both planetesimals and dust grains, but not pebbles. For this work we model KELT-9b assuming the planet initiated its migration from distances between $350$\,AU and $0.0355$\,AU, the current position of KELT-9b from its host star \citep{Ahlers2020, Wong2021b}. By requiring that the fully assembled planet has a mass equal to that of  KELT-9b, we can estimate the mass of the heavy elements KELT-9b can have accreted in the form of planetesimals during its type II migration. This mass depends on two parameters: KELT-9b's initial orbital distance and the disk's planetesimal ratio, which defines the presence of planetesimals in the disk as well as how efficient a planet is at accreting planetesimals.

SimAb shows that for a planet like KELT-9b to reach a metallicity of  100$\times$solar, it must have initiated its type II migration at approximately $305$\,AU. In this case KELT-9b must accrete all the planetesimals available in the disk. However a capture fraction of 100\%\  is not realistic \citep{Shibata2020}. This means the planet must have initiated its type II migration even farther away than this distance. For example, if we assume the planet on average accretes half of the available planetesimals, this distance would be as large as $500$\,AU. These distances are unrealistically large. Moreover this is only possible if we assume an alpha parameter very close to one, while the observed values suggest the alpha parameter to be much less than that \citep{Pinte2016, Flaherty2018}.
In addition, given its high spin-orbit angle \citep{Ahlers2020}, KELT-9b is unlikely to have had type II migration all the way to its current location, and hence most likely did not accrete as many planetesimals as calculated above.

It is notable that under the assumptions made in SimAb,  KELT-9b has likely accreted very little matter within the water-ice line, matter that can significantly lower the C/O ratio. Yet, we retrieve a very low C/O ratio from our spectrum, even lower than SimAb can produce. In our high-metallicity scenario, it is therefore necessary that KELT-9b has accreted a large portion of its planetesimals while it is within the water-ice line, hence it should have undergone late-stage pollution in a high-metallicity scenario.

\end{document}